\pdfoutput=1  
\documentclass[aps,prb,twocolumn,nobibnotes]{revtex4-2}
\usepackage{dcolumn}
\usepackage{graphicx}
\usepackage{color}
\usepackage{amssymb}
\usepackage{amsmath}
\usepackage{bm}
\usepackage[colorlinks=true,linktocpage=true,breaklinks=true]{hyperref}
\usepackage{multirow}
\usepackage{braket}
\usepackage{appendix}

\begin{document}
\title{Phonon-Mediated Superconductivity near Phase Boundaries in Rhombohedral Multilayer Graphene}
\author{Konstantinos Sourounis}
\email{konstantinos.sourounis@physics.uu.se}
\author{Peter M. Oppeneer}
\email{peter.oppeneer@physics.uu.se}
\affiliation{Department of Physics and Astronomy, Uppsala University, P.\ O.\ Box 516, SE-75120 Uppsala, Sweden}

\begin{abstract}
The recent observation of superconductivity in untwisted multilayer graphene has sparked intense interest in its microscopic origin. As the displacement field and carrier density are varied, superconductivity emerges in several distinct regions with different temperature and magnetic-field responses. This behavior has been attributed to the presence of distinct electronic phases, raising the possibility that different superconducting regions originate from different microscopic mechanisms. While purely electronic pairing mechanisms have been extensively explored,  an in-depth examination of phonon-mediated superconductivity is still missing. Here, we solve the coupled selfconsistent Migdal-Eliashberg equations using input from first-principles calculations and incorporating interaction-driven symmetry breaking of the electronic ground state. Varying the carrier density and displacement field, we reproduce the experimentally observed superconducting regions with comparable superconducting properties such as gap and $T_c$ and predict additional, as-yet undiscovered superconducting phases. Our findings exemplify that 
several superconductivity regions in rhombohedral graphene can be satisfactorily explained by phonon-mediated pairing. 
\end{abstract}

\maketitle
{\textit{Introduction}} - After the discovery of superconductivity in twisted multilayer graphene \cite{cao2018unconventional, park2021tunable, park2022robust}, superconductivity has been discovered in a variety of crystalline rhombohedral graphene systems including two (2G) \cite{Zhou2022Bilayer,zhang2023enhanced,li2024tunable}, three (3G) \cite{Zhou2021Trilayer,patterson2024superconductivity,yang2025impact}, four (4G), and five (5G) \cite{Han2025tetralayer,seo2025family,Choi2025,zhou2026competing}, or more layers (NG) \cite{deng2025superconductivity,guo2025flat,kumar2025pervasive,kumar2025superconductivity,xie2026magnetic,Zheng_2026,nguyen2025hierarchy,qin2025extreme,nguyen2026coexisting,hua2026multi}. Moiré potentials, spin-orbit coupling, and magnetic fields can induce or suppress superconductivity, whose critical temperature generally remains well below $T_c\leq 500$\,mK \cite{Zhou2021Trilayer,Han2025tetralayer,guo2025flat,kumar2025superconductivity}. Increasing the number of layers leads to a higher density of states (DoS) at the Fermi level, but the superconducting $T_c$ has not been observed to increase similarly. Unlike twisted multilayer graphene, rhombohedral graphene exhibits superconductivity even in the absence of a Moir{\'e} potential. Moreover, the electronic ground state and magnetic-field response indicate the possibility of diverse underlying mechanisms for superconductivity. Numerous theoretical proposals based on purely electronic mechanisms have been put forward, including Coulomb interaction \cite{cea2022superconductivity,jimeno2023superconductivity,geier2025chiral,parra2025band}, Kohn-Luttinger mechanism \cite{ghazaryan2021unconvnetional,you2022kohn,ghazaryan2023multilayer,chou2025intravalley}, spin fluctuations \cite{chatterjee2022inter,dong2023transformer,raines2026superconductivity,dong2026superconductivity}, quantum geometric origin \cite{jiang2026ideal,jahin2026enhanced,shavit2025quantum}, and other electronic mechanisms \cite{kumar2025unconvnetional,chiho2026quarter}. 

Phonon-mediated superconductivity in rhombohedral graphene has so far been studied only to a limited extent \cite{chou2021acoustic,chou2022acoustic1,chou2022acoustic2,li2023charge,vinas2024phonon}, with low-energy acoustic phonons being identified as the  likely pairing modes. Estimates of the electron-phonon coupling in most studies were however mostly phenomenological, and the symmetry breaking due to electron-electron interaction, which appears essential for reproducing the experimental phase diagram, was not included. Experimental studies indicate that electron-phonon coupling in graphene can be strong, \cite{yan2007electric,efetov2010controlling,zan2024electron,chen2024strong}, consistent with first-principles calculations \cite{yan2008phonon,yan2009electron,park2008electron,borysenko2010first,borysenko2011electron,gao2024first}. A fully selfconsistent treatment that accounts for electron-phonon retardation, an important effect established since the early days of superconductivity, is still lacking for multilayer graphene. Migdal-Eliashberg theory provides an accurate quantitative framework for describing phonon-mediated superconductivity while accounting for retardation effects \cite{margine2013,aperis2015mgb2,giustino2017,aperis2018multiband,sanna2020combining,Das2023,marini2023phonon,mishra2025electron}. 

In this work, we use first-principles electron and phonon calculations to investigate the different superconducting regions as the number of layers varies. Combining interaction-driven symmetry breaking with \textit{ab initio} electron-phonon coupling, we solve selfconsistently the coupled Migdal-Eliashberg equations to reveal superconductivity across a range of displacement fields and carrier densities, obtaining results consistent with experiments. Additionally, we predict superconductivity in as-yet unexplored regions. Finally, we discuss whether the experimentally observed superconducting regions are more naturally explained by electron-phonon pairing or by purely electronic mechanisms.

{\it{Electronic Structure \& Symmetry Breaking}} --
In the absence of interactions, graphene possesses four degenerate flavors, $\alpha=\{1,2,3,4\}=\{{\rm K}\!\uparrow, {\rm K}'\!\uparrow, {\rm K}\!\downarrow, {\rm K}'\!\downarrow\}$, corresponding to the valley and spin degrees of freedom. Depending on the carrier density and applied displacement field, multilayer rhombohedral graphene can exhibit distinct phases, the fourfold-degenerate metallic ($\gamma = 4$), half-metallic ($\gamma = 2$), or quarter-metallic ($\gamma = 1$) phases. The experimental discovery of superconductivity, first in trilayer graphene \cite{Zhou2021Trilayer,zhou2021half}, revealed that superconductivity emerges near transitions between these phases. Experiments \cite{sharpe2019emergent,zhou2021half, Zhou2022Bilayer,de2022cascade,kalantre2026fermiology} and theory \cite{ghazaryan2023multilayer,xie2023flavor,koh2024correlated,wang2024electrical,david2025valley,parra2025nematic,miao2026various} have consistently reported symmetry breaking occurring between these flavors in association with higher DoS, thereby favoring superconductivity near the phase boundaries separating different flavor-polarized states.

We describe the electronic structure of rhombohedral multilayer graphene using a tight-binding model parameterized from first-principles calculations \cite{koshino2009trigonal,zhang2010band,xiao2011density,parra2025nematic,kumar2025superconductivity}. The model captures the evolution of the band structure with increasing layer number. 
In the absence of the displacement field, the conduction and valence bands touch at the K and K$'$ points. Applying a displacement field opens a gap at the K and K$'$ points and modifies the Fermi-surface topology, see Supplementary Materials (SM) \cite{SM}. Together with carrier doping, the displacement field provides an effective means of tuning the Fermi surface and, consequently, the electronic DoS $\rho$. The details of the Hamiltonian can be found in the SM \cite{SM}. Increasing the number of graphene layers also enlarges the Fermi surface, generally leading to a higher electronic DoS. Our calculations reproduce the experimentally observed evolution of the DoS: with increasing layer number, the hole-side DoS shifts toward lower displacement fields, whereas the electron-side DoS shifts toward higher fields. These systematic changes in the electronic DoS play a central role in the symmetry-breaking, electronic screening, and superconductivity discussed below.

Electron-electron interactions lift the fourfold flavor degeneracy and generate the sequence of symmetry-broken phases observed experimentally. To describe this behavior, we employ the rigid-band Stoner approximation \cite{zondiner2020cascade,zhou2021half,ghazaryan2023multilayer} to model the symmetry breaking between different flavors. 
The grand potential for our problem is
\begin{eqnarray}
   \!\! \! \!\frac{\Phi}{A_{\rm uc}}\!\!&=&\!\!\sum_{\alpha}\big\{ E_0(n_{\alpha})-\mu_{\alpha}n_{\alpha}\big\}\nonumber\\
    &+&\! \!\frac{UA_{\rm uc}}{2}\sum_{\alpha\neq \beta}n_{\alpha}n_{\beta}+\frac{JA_{\rm uc}}{2}(n_1-n_3)(n_2-n_4) ,
\end{eqnarray}
 where $U$ denotes the Coulomb interaction and $J$ the Hund's coupling. $E_0 (n_{\alpha})$ is the kinetic energy, $n_{\alpha}$ the charge density for flavor $\alpha$,  and $\mu_{\alpha}$ is the corresponding  chemical potential. Throughout this work, we consider only ferromagnetic Hund's coupling ($J<0$). The unit-cell area is $A_{\rm uc}=\sqrt{3}a^2/2$. Since $U$ and $J$ are not known \textit{apriori}, we treat them as phenomenological parameters chosen to reproduce experiment and allow them to vary with the number of layers. The noninteracting DoS ($U=J=0$) and results obtained for several parameter sets for the $N=4$ and $N=5$ cases are presented in the SM \cite{SM}.
 
\begin{figure}[t]
\includegraphics[width=8.6cm,height=8.6cm,keepaspectratio]{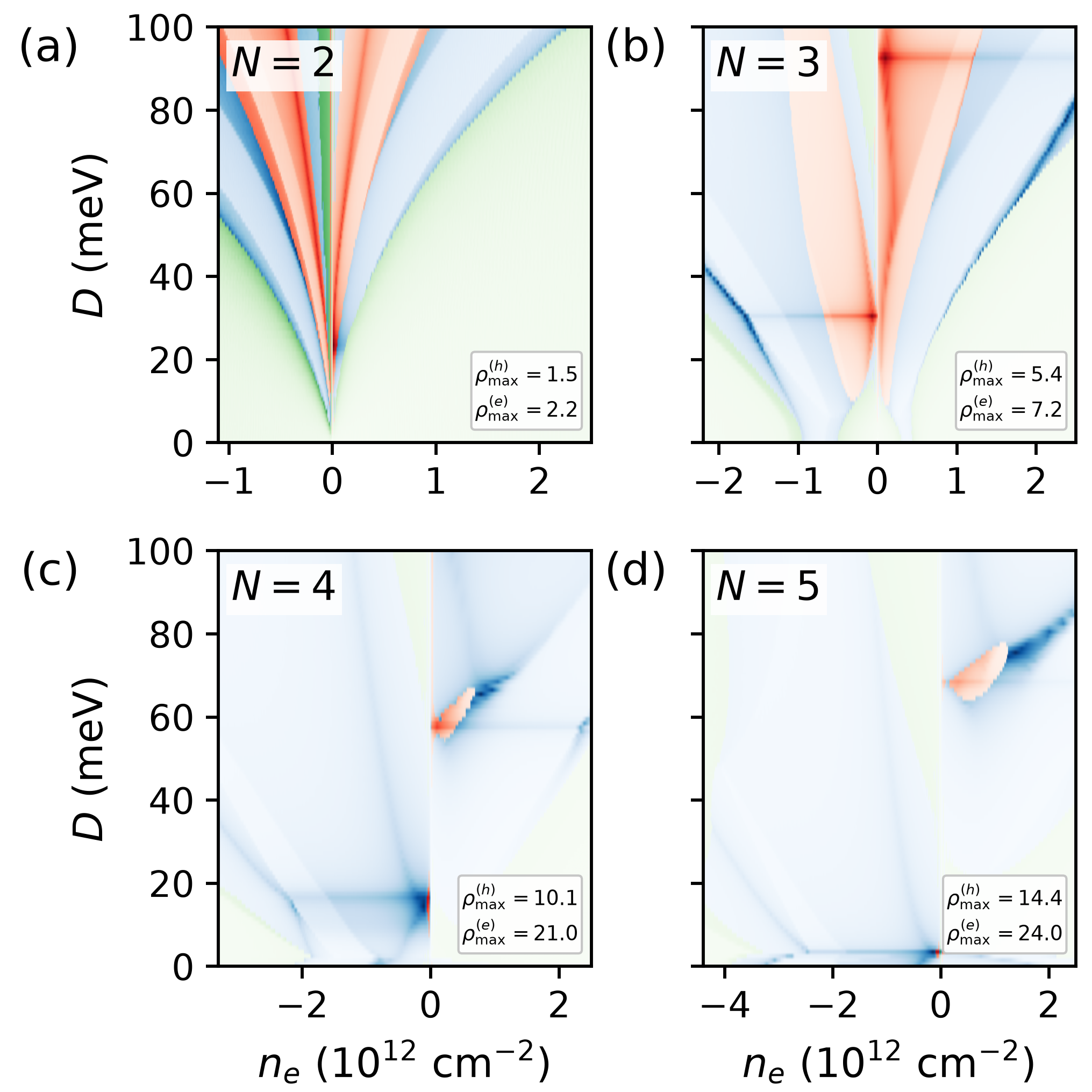}
\vspace*{-0.5cm}
\caption{DoS calculated  as function of electron doping $n_e$ and displacement field $D$ for rhombohedral graphene with (a)-(d): $N=2,\, 3,\, 4,$ and 5 layers. Colors denote the flavor degeneracy: full metal ($\gamma = 4$, green), half-metal ($\gamma = 2$,
blue), and quarter-metal ($\gamma = 1$, red). The colormap is normalized to the maximum density of states in the electron- and hole-doped region, with $\rho_{\rm max}^{(h)}$
and $\rho_{\rm max}^{(e)}$ given in each panel. Units are in $\rm{eV^{-1}nm^{-2}}$. The Stoner parameters are $(U, J) = (60, -9)$, $ (30, -9)$, $(13, -18)$ and $(11, -14)$ eV for (a)-(d).}
\label{fig:2}
\end{figure}

Minimizing the grand potential yields the interaction-induced symmetry-broken phases; their calculated DoS as function of electron doping and displacement field is shown in Fig.\ \ref{fig:2}. The largest DoS occurs near the boundaries between competing symmetry-broken phases, precisely where superconductivity is observed experimentally. This demonstrates that electron-electron interactions are essential for reproducing both the experimentally observed flavor symmetry breaking and the high-DoS regions favorable for superconductivity. The rigid-band Stoner approach is an approximation but it provides a qualitative description of the symmetry breaking.  

\begin{figure}[tb]
\includegraphics[width=8.6cm,height=8.6cm,keepaspectratio]{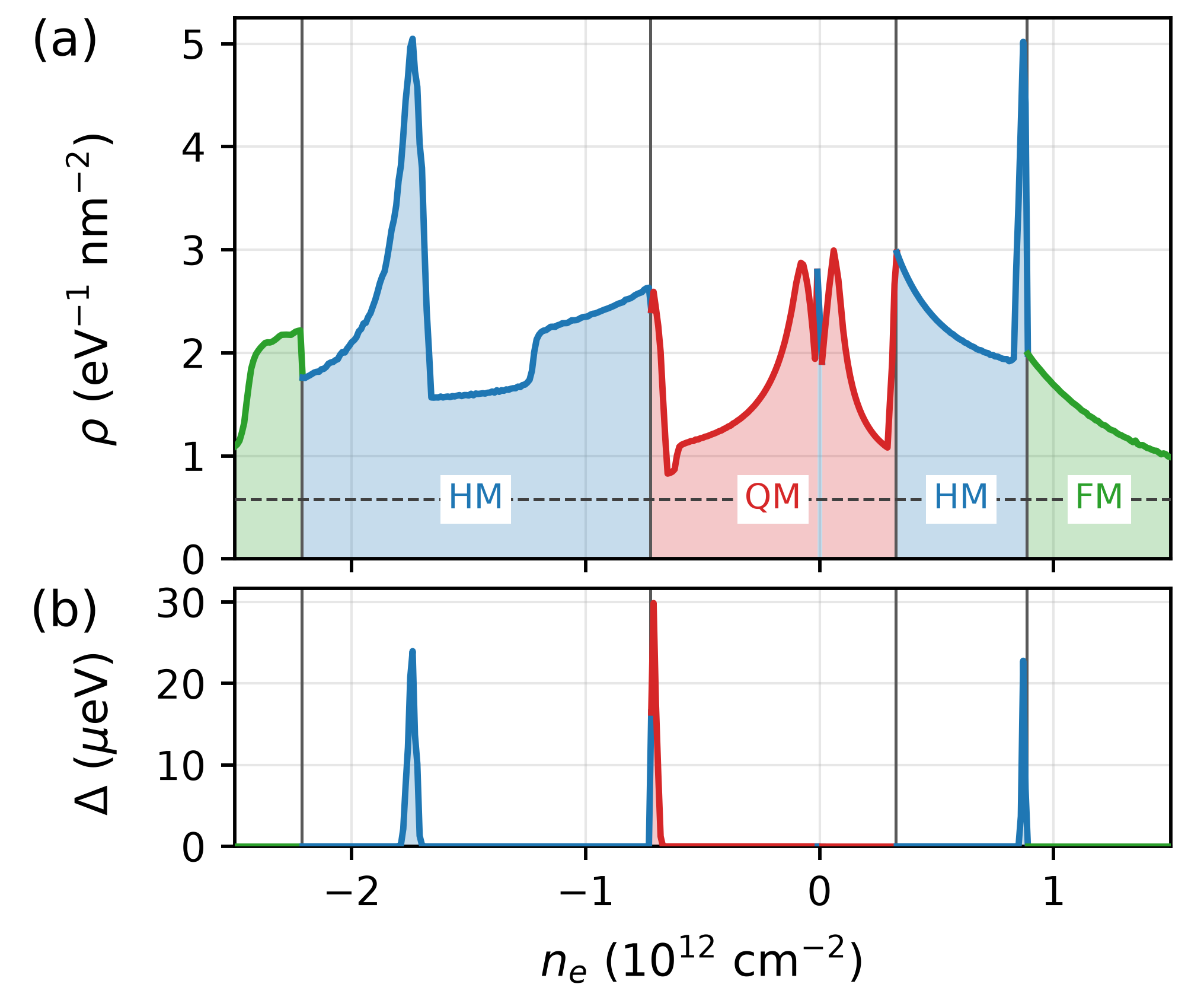}
\vspace*{-0.7cm}
\caption{(a) Interaction-enhanced density of states $\rho$ of rhombohedral trilayer graphene  as function of electron doping and 
at displacement $D = 32$ meV. Colors denote the flavor degeneracy: full metal ($\gamma = 4$, green), half-metal ($\gamma = 2$, blue) and quarter-metal ($\gamma = 1$, red). 
(b) Superconducting gap obtained from the selfconsistent Migdal-Eliashberg equations at the same densities and $T = 25$\,mK.}
\label{Fig:1}
\end{figure}

Another consequence of the interaction is the sequence of Stoner-like transitions from the fourfold-degenerate metal to the half-metal and finally the quarter-metal as the DoS increases. An example of these transitions is shown for trilayer graphene in Fig.\ \ref{Fig:1}(a): sizable variations of the DoS occur near Stoner transitions between different flavors. For comparison, we show the selfconsistently calculated superconducting gap $\Delta$ in Fig.\ \ref{Fig:1}(b) (details provided further below). Notably, superconductivity emerges in high-DoS regions, but is not the only determining quantity.

Because the available states become concentrated into fewer flavors, superconductivity must be treated within the corresponding flavor sector. Accordingly, we define the DoS per flavor as $N(E_{\rm F})=\rho/g$, where $g$ denotes the flavor degeneracy. This choice is consistent with transport experiments \cite{zhou2021half}, which observe an approximately twofold reduction in resistance across the transition from the fourfold-degenerate metal to the half-metal, indicating that the DoS per flavor remains approximately unchanged across the transition.

\begin{figure}[t]
\includegraphics[width=8.0cm,height=8.0cm,keepaspectratio]{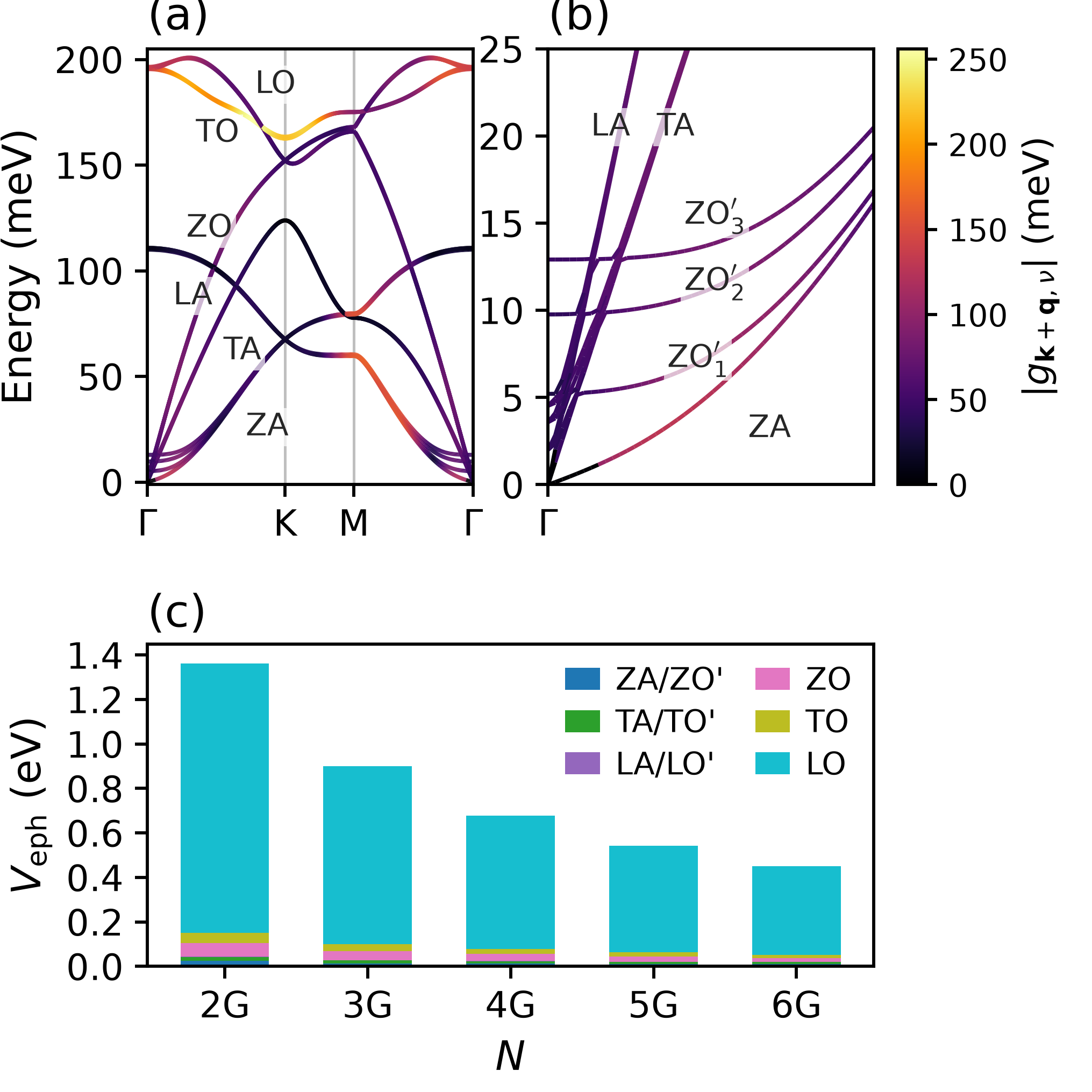}
\vspace*{-0.5cm}
\caption{(a) First-principles phonon dispersion of rhombohedral tetralayer graphene along the high-symmetry path of the Brillouin zone. (b) Enlargement of the phonon dispersion near the $\Gamma$ point. The color scale indicates the magnitude of the electron-phonon matrix elements $g_{\bm{k}+\bm{q},\nu}$ shown in the colorbar. (c) Phonon-branch resolved electron-phonon coupling $V_{eph}$ versus the number of graphene layers, for scattering of electrons at the K valley. The bar height gives the total electron-phonon coupling.}
\label{fig:3}
\end{figure}

{\textit{Electron-Phonon Interaction}} -- While phonons and electron-phonon interactions have been investigated in graphene with up to three layers \cite{yan2008phonon,yan2009electron,park2008electron,borysenko2010first,borysenko2011electron,gao2024first}, corresponding studies of thicker rhombohedral multilayers remain limited. Here, we use Wannier-interpolated density functional theory (DFT) and density functional perturbation theory (DFPT) \cite{giannozzi2009quantum,giannozzi2017advanced,pizzi2020wannier90,zhou2021perturbo} to calculate the evolution of the phonon dispersion and electron-phonon coupling with increasing layer number. As the number of layers increases, the overall phonon dispersion remains largely unchanged because of the weak interlayer coupling. As in monolayer graphene, the phonon spectrum consists of out-of-plane, transverse, and longitudinal optical (ZO, TO, LO) and acoustic (ZA, TA, LA) branches. The primary differences emerge near the $\Gamma$ point, where additional modes appear, most notably the breathing (ZO$'$) and shear (TO$'$/LO$'$) modes. Figures \ref{fig:3}(a) and (b) show the first-principles phonon dispersion of tetralayer graphene across the Brillouin zone and near the $\Gamma$ point. Details of the calculations, together with the phonon dispersions for $N=2,\,3,\,5,$ and 6, are provided in the SM \cite{SM}.

Although the phonon spectrum is largely insensitive to the number of layers, the electron-phonon coupling decreases substantially, indicating that the layer dependence originates primarily from the electronic rather than the phononic sector. From the DFT+DFPT calculations we extract the phonon frequencies $\omega_{q\nu}$ and electron-phonon matrix elements $g_{\bm{k}+\bm{q},\nu}$ using Wannier interpolation \cite{zhou2021perturbo}. We then compute the mode-resolved and total electron-phonon coupling at the K valley,
\begin{eqnarray}
\lambda^0_{\bm{q}\nu}=2\frac{N(0)|g_{\bm{k}+\bm{q},\nu}|^2}{\hbar\omega_{\bm{q}\nu}} ~~ \textrm{and} ~~ V_{eph}=\frac{1}{N}\sum_{\bm{q},\nu}\lambda^0_{\bm{q}\nu} .
\end{eqnarray}
Here, we normalize the electronic DoS by setting $N(0)=1$. The resulting electron-phonon coupling $V_{eph}$ is shown in Fig.\ \ref{fig:3}(c). The electron-phonon coupling decreases systematically with increasing layer number $N$. Specifically, the total coupling strength in hexalayer graphene is approximately half that in trilayer graphene. We attribute this reduction to enhanced electronic screening associated with the increasing Fermi-surface area as the number of layers grows.

\begin{figure}[t]
\includegraphics[width=8.6cm,height=8.6cm,keepaspectratio]{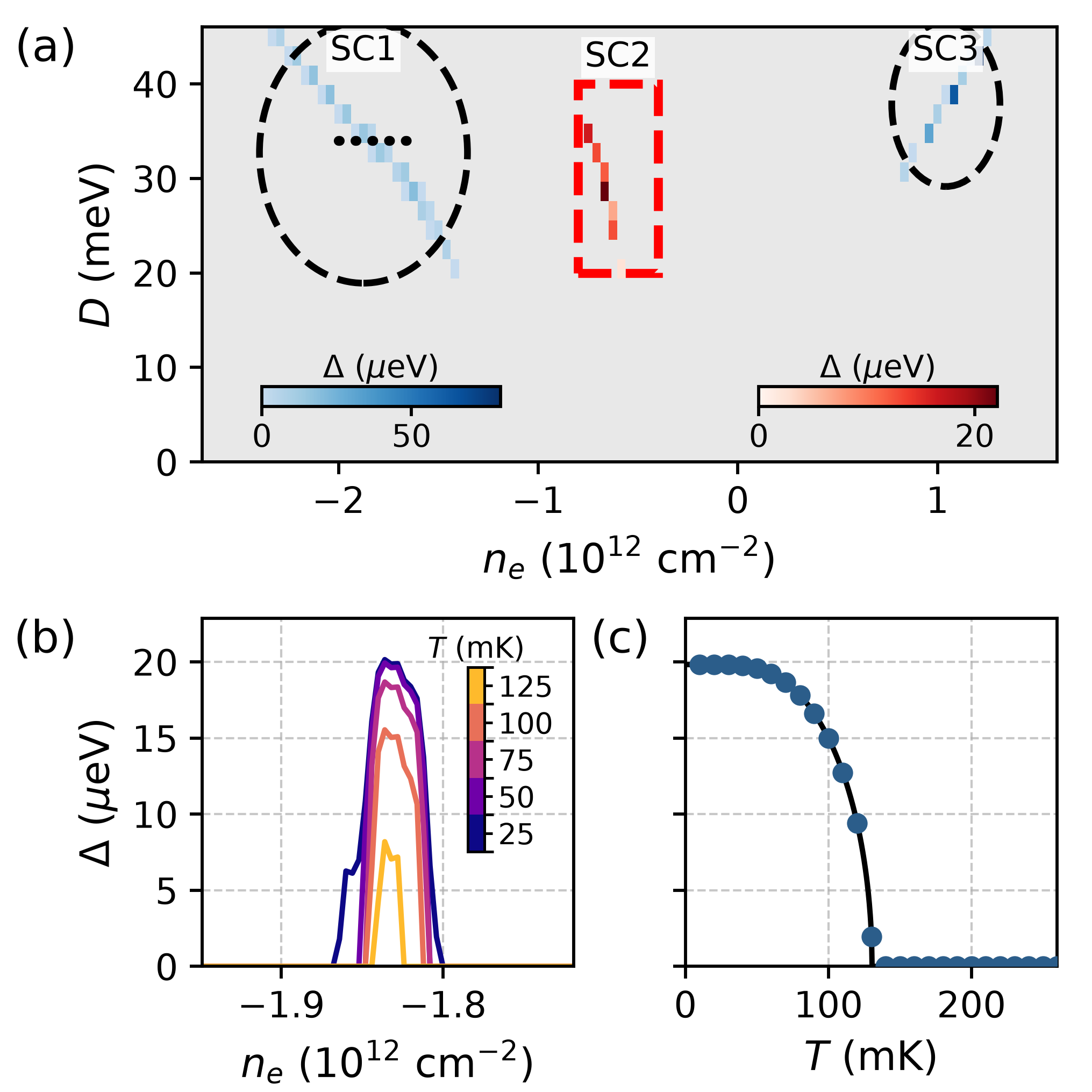}
\vspace*{-0.6cm}
\caption{Superconductivity in rhombohedral trilayer graphene.
(a) Calculated superconducting gap, colored by flavor degeneracy: $\gamma = 2$ (blue, SC1 and SC3) and $\gamma = 1$ (red, SC2); the two inset color scales give $\Delta$ in $\rm{\mu}$eV, blue for $\gamma = 2$ and red for $\gamma = 1$. The red dashed rectangle marks the $\gamma = 1$ window corresponding to SC2. The calculations are done for $T = 75$\,mK. (b) Superconducting gap evaluated along the gray dotted line in panel (a).
(c) Temperature dependence of the gap at the maximum of panel (b).}
\label{fig:4}
\end{figure}

{\it{Superconductivity}} -- The superconducting phase diagram emerges from the interplay between the interaction-enhanced electronic DoS and the layer-dependent electron-phonon coupling. To incorporate the interaction-enhanced electronic DoS, we rescale the electron-phonon coupling according to $\lambda^{E_{\rm F}}_{\bm{q}\nu}=N(E_{\rm F})\lambda^0_{\rm{q}\nu}$, where $N(E_{\rm F})$ is obtained from the interaction calculations shown in Fig.\ \ref{fig:2}.  To determine the resulting superconducting phase diagram, we solve the coupled selfconsistent Migdal-Eliashberg equations  using the Uppsala Superconductivity Code \cite{aperis2015mgb2,aperis2018multiband,schrodi2019fesesto,schrodi2020prominet,Schrodi2024}, see the SM \cite{SM} for details.

We first consider trilayer graphene, where experiments have identified two distinct superconducting regions in the hole-doped regime, denoted SC1 and SC2 \cite{Zhou2021Trilayer}. SC1 appears at the phase boundary between the full- and half-metal and is suppressed by an applied magnetic field, consistent with conventional electron-phonon superconductivity. Solving the Migdal-Eliashberg equations for the $\gamma = 4$ and $\gamma = 2$ phases fully reproduces the experimentally observed SC1 region, see Fig.\ \ref{fig:4}(a). Figure \ref{fig:4}(b) shows the calculated temperature dependence of the superconducting gap, demonstrating very good quantitative agreement with measurements (cf.\ Fig.\ 1(h) in Ref.~\cite{Zhou2021Trilayer}). Additionally, we predict a yet-to-be-detected region in the electron-doped region arising between the full- and half-metal regions, with similarities to SC1 in the hole-doped region, which we denote as SC3. 
SC2 is observed at the boundary between the half-metal and quarter-metal phases and, unlike SC1, remains robust under strong in-plane magnetic fields \cite{Zhou2021Trilayer}. Solving the Migdal-Eliashberg equations in this regime likewise yields a superconducting state localized near the half-metal/quarter-metal phase boundary [Fig.~\ref{fig:4}(a)], although the magnetic-field response of SC2 might suggest involvement of an unconventional pairing mechanism
\cite{arp2024intervalley}.  We note that additional superconducting regions are found in our solution near \(n_e = 0\). However, in experiments, these regions are insulating due to a small gap induced by the presence of h-BN \cite{giovanneti2007,hunt2012}. The unfiltered data are presented in the SM \cite{SM}. The calculated SC1 gap value, \(\Delta \approx 20\,\mu\mathrm{eV}\), and $T_c \approx 125$\,mK are consistent with the BCS relation \(\Delta \approx 1.764 \,k_{\rm B} T_c\). These values are furthermore close to the experimentally observed \(T_c \approx 120-150\,\mathrm{mK}\) \cite{Zhou2021Trilayer}. This agreement supports electron–phonon coupling as a plausible mechanism for the observed superconductivity in trilayer graphene. We note that these results are obtained for the Coulomb pseudopotential $\mu^* = 0$; including $\mu^*$ derived from many-body calculations is discussed further below.

\begin{figure}[t]
\includegraphics[width=8.6cm,height=8.6cm,keepaspectratio]{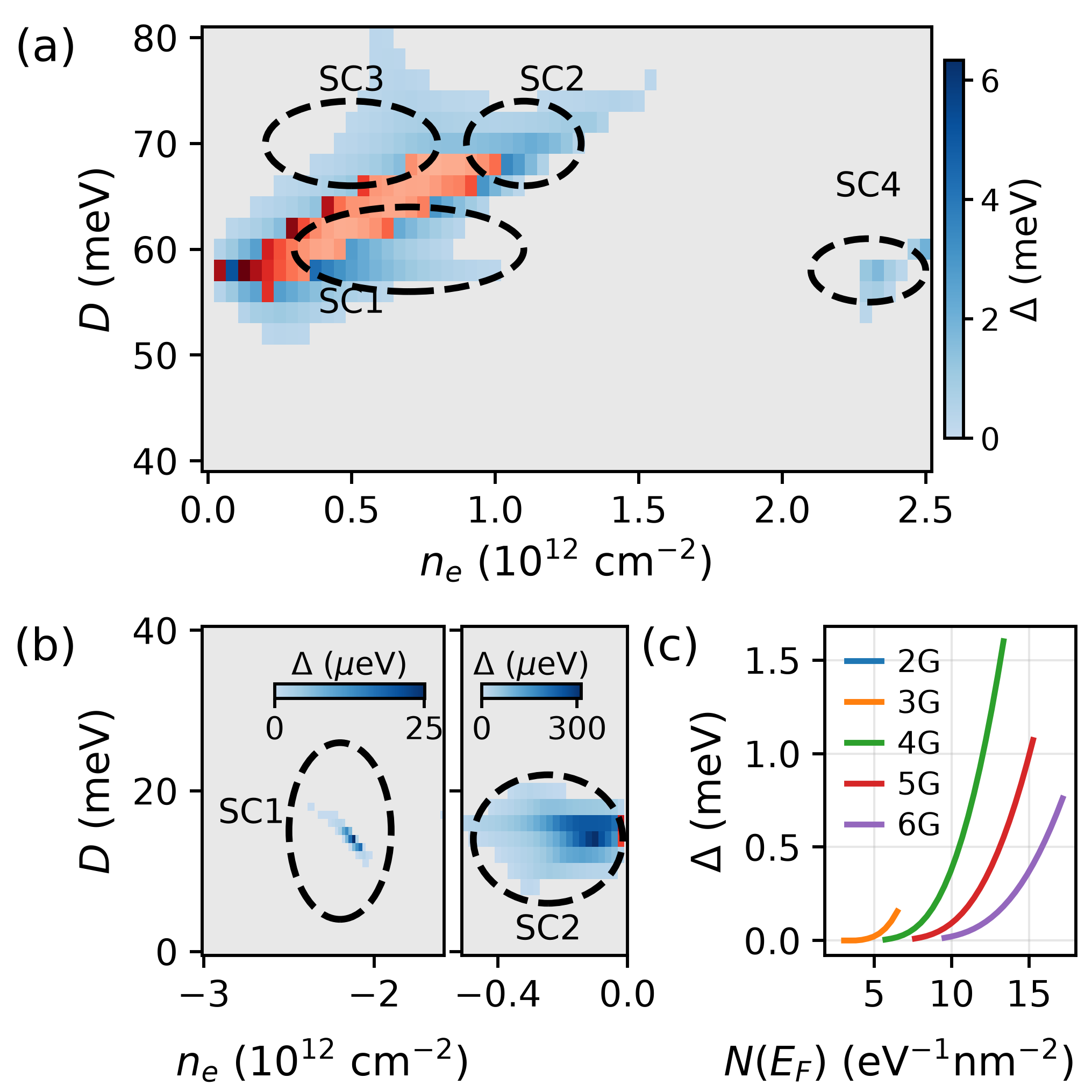}
\vspace*{-0.6cm}
\caption{Predicted superconductivity in rhombohedral tetra-layer graphene. (a) Appearance of various superconducting gaps (SC1 $-$ SC4) in the electron-doped regime. (b) Superconducting gap in the hole-doped regime  with two superconducting regions SC1 and SC2. (c) Maximum superconducting gap as a function of the density of states obtained using the first-principles electron-phonon coupling for different numbers of graphene layers. All calculations are done for $T=25$\,mK.}
\label{fig:5}
\end{figure}

We next examine how the superconducting phase diagram evolves with increasing layer number. For tetralayer graphene, we find several superconducting regions in the electron- and hole-doped regimes (Figs.\ \ref{fig:5}(a) and \ref{fig:5}(b), respectively). In the hole-doped regime we obtain an extended superconducting region [Fig.~\ref{fig:5}(b)], similar to the SC1 region of the trilayer; this region has recently been experimentally reported \cite{seo2025family}. Additionally, experiments identify a second superconducting region, SC2, at lower dopings, characterized by a lower critical temperature $T_c$ \cite{seo2025family}. Our theory also predicts the emergence of this second region. However, the calculated gap is substantially larger than would be expected from the experimental critical temperature of $T_c \approx 50$ mK. We attribute this discrepancy to the presence of the h-BN-induced gap near charge neutrality, $n_e \approx 0$, in experiments. The electron-doped regime, shown in Fig.\ \ref{fig:5}(a), exhibits substantially larger superconducting gaps. At low electron doping, superconductivity extends over a broad region, which separates into three phases SC1-SC3 that have been experimentally observed recently \cite{Han2025tetralayer}. SC1 and SC2 exhibit the largest superconducting gaps owing to the enhanced DoS in these regions. Their experimentally observed robustness against out-of-plane magnetic fields may, however, points to the involvement of other pairing channels or to a more exotic pairing mechanism. The SC3 region, which is well developed in tetralayer graphene, becomes progressively weaker in pentalayer graphene \cite{Han2025tetralayer}, and is no longer observed in hexalayer graphene \cite{nguyen2025hierarchy,qin2025extreme}. Our calculations also predict an as-yet undiscovered superconducting phase, denoted SC4, at the half-metal/full-metal phase boundary, analogous to the SC3 region red predicted for trilayer graphene.

The evolution of superconductivity with layer number is governed by the competition between the decreasing electron-phonon coupling and the increasing electronic DoS. To quantify this competition, we solve the Migdal-Eliashberg equations using the first-principles electron-phonon coupling for each layer together with the corresponding interaction-determined range of DoS. Figure~\ref{fig:5}(c) summarizes the resulting superconducting gap over a range of DoS for different layer numbers. As the number of layers increases, progressively larger DoS are required to sustain superconductivity. This requirement competes with the interaction-driven Stoner transitions, naturally confining superconductivity to the phase boundaries where the DoS is maximized. For untwisted bilayer graphene ($N=2$), our calculations predict absence of phonon-mediated superconductivity, consistent with experiment \cite{Zhou2022Bilayer}. For pentalayer and hexalayer graphene, even larger densities of states are required to sustain superconductivity, although we find that the overall phase diagram remains qualitatively similar to that of tetralayer graphene.

{\it{Discussion}} - By combining first-principles electron-phonon interactions with interaction-driven symmetry breaking, we present the first systematic study of phonon-mediated superconductivity across rhombohedral graphene multilayers. Comparison with experiment shows excellent agreement for trilayer graphene, whereas for tetralayer graphene the observed superconducting regions are correctly predicted, but the superconducting gap is overestimated.  
An important outcome of this work is that electron-electron interactions play a crucial role even when they are not the sole pairing mechanism for superconductivity. Furthermore, we find that accurately determining the symmetry-broken ground state and the corresponding electronic DoS is essential for obtaining the correct superconducting gap magnitude. 
 
We mention here three key aspects that may further improve the quantitative agreement. First, a fully selfconsistent Hartree-Fock calculation \cite{parra2025band,parra2025nematic,guo2025flat,wang2024electrical,koh2024correlated} of the interaction-enhanced DoS would provide a more accurate description of the symmetry-broken phases. Second, the present treatment neglects the electronic self-energy arising from electron-electron interactions, which is expected to broaden the sharp density-of-states 
 peaks produced by the rigid-band approximation.
 Third, it is common for solutions of the Migdal-Eliashberg equations to include the Anderson-Morel Coulomb pseudopotential \(\mu^*\), to account for 
the Coulomb repulsion that suppresses phonon-mediated superconductivity \cite{margine2013,giustino2017,vinas2024phonon,Das2023}. The usual approach, in which \(\mu^*\) is chosen as a standard value, is not applicable to graphene multilayers because the DoS changes substantially with displacement field and doping, necessitating a more sophisticated treatment. However, determining the value of \(\mu^*\) from many-body interactions remains an open problem \cite{Schonhoff2016,simonato2023,wang2023,spataru2026,cai2025superconductivity}. In the SM \cite{SM}, we present a calculation of the Migdal–Eliashberg equations including \(\mu^*\) determined from many-body calculations. We distinguish two pairing regimes. In the intra-valley channel, small-momentum Coulomb interactions can promote pairing via the Kohn–Luttinger mechanism \cite{ghazaryan2021unconvnetional,you2022kohn,parra2025band,geier2025chiral}, motivating $\mu^*\leq 0$.  
In the inter-valley channel, however, incorporating $\mu^*>0$ 
renormalizes large bare gaps closer to experimentally observed energy scales.

The present work focuses on conventional s-wave phonon-mediated superconductivity, which provides a natural explanation for the superconducting regions suppressed by either in-plane or out-of-plane magnetic fields. However, several theoretical studies have shown that electron-phonon interactions can also stabilize unconventional superconducting states
\cite{schnell2006unconventional,brydon2014odd,schrodi2021unconvnentional,kristoffer2025phonon,vinas2024phonon,chou2025superconductivity}. In particular, superconducting regions that remain robust against strong in-plane magnetic fields, such as SC2 in trilayer graphene, are consistent with an Ising-like superconducting state in which spin-orbit coupling locks the electron spins perpendicular to the graphene plane, protecting Cooper pairs against in-plane Zeeman fields beyond the Pauli limit \cite{arp2024intervalley,Das2023}. Finally, the relatively strong electron-phonon coupling found in these systems suggests that nonadiabatic effects beyond the Migdal approximation may become important. Such effects have already been identified in monolayer and bilayer graphene, and extending their investigation to rhombohedral multilayers represents a promising direction for future work \cite{nonadiabatic2025}. 
While our results show that electron-phonon interactions naturally 
explain parts of the superconducting phase diagram, other superconducting regions may involve additional electronic 
pairing mechanisms. We propose that experiments with increasing screening, as already performed in twisted bilayer graphene \cite{Stepanov2020,Saito2020,Liu2021Tuning,prx2026coulomb}, could shed light on the nature of the pairing mechanism. If superconductivity persists under stronger screening, this would support an electronic pairing mechanism, whereas its suppression with increasing screening would provide evidence for a phonon-mediated pairing mechanism.

\textit{Conclusions -}
Our calculations reproduce all observed superconducting regions in trilayer and tetralayer graphene both for electron- and hole-doping. Excellent agreement between experiment and theory is found for the SC1 region in the hole-doped regime, first observed in trilayer graphene \cite{Zhou2021Trilayer} and recently reported in tetralayer and pentalayer devices \cite{seo2025family}.  In addition, we predict a family of new superconducting regions, SC3 in trilayer and SC4 in tetralayer, in the electron-doped regime at the boundary between the full-metal ($\gamma = 4$) and half-metal ($\gamma = 2$) phases. Lastly, we find that the systematic reduction of the electron-phonon coupling with increasing layer number provides a natural explanation for why the superconducting $T_c$ has not been observed to increase despite the growing electronic density of states. Our work establishes a quantitative framework that combines correlated electronic phases with first-principles electron-phonon calculations and can readily be extended to other two-dimensional multilayer systems, including superconducting transition-metal dichalcogenides. 

\textit{Acknowledgments} --
K.S.\ thanks Alejandro Jimeno-Pozo, Markus Weißenhofer and Chandan K.\ Singh for fruitful discussions. This work was supported by the Swedish Research Council (VR) and the Knut and Alice Wallenberg Foundation (Grants No.\ 2022.0079 and No.\ 2023.0336). The calculations were supported by resources provided by the National Academic Infrastructure for Supercomputing in Sweden (NAISS) at NSC Linköping, partially funded by VR through Grant No.\ 2022-06725.

\bibliography{bibliography}

\clearpage
\onecolumngrid   

\setcounter{page}{1}
\setcounter{equation}{0}
\setcounter{figure}{0}
\setcounter{table}{0}
\setcounter{section}{0}
\renewcommand{\thepage}{S\arabic{page}}
\renewcommand{\theequation}{S\arabic{equation}}
\renewcommand{\thefigure}{S\arabic{figure}}
\renewcommand{\thetable}{S\arabic{table}}
\renewcommand{\theHequation}{S\arabic{equation}}
\renewcommand{\theHfigure}{S\arabic{figure}}
\renewcommand{\theHtable}{S\arabic{table}}
\renewcommand{\theHsection}{SM\arabic{section}}

\begin{center}
  {\small\textsc{Supplemental Material}\par}
  \vspace{8pt}
  {\large\bfseries Phonon-Mediated Superconductivity near Phase Boundaries in Rhombohedral\\[2pt]
  Multilayer Graphene\par}
  \vspace{10pt}
  Konstantinos Sourounis and Peter M.\ Oppeneer\par
  \vspace{2pt}
  {\small\itshape Department of Physics and Astronomy, Uppsala University, P.\ O.\ Box 516, SE-75120 Uppsala, Sweden\par}
\end{center}
\vspace{12pt}

This Supplemental Material documents the calculations behind the main text: the tight-binding model (Sec.~\ref{sec:tb}), the rigid-band Stoner treatment of the flavor symmetry breaking (Sec.~\ref{sec:stoner}), the first-principles phonons and electron-phonon matrix elements (Sec.~\ref{sec:dft}), the numerical solution of the Migdal-Eliashberg equations (Sec.~\ref{sec:me}), and the effect of the Coulomb pseudopotential (Sec.~\ref{sec:mustar}).  Figures, tables and equations of the main text are referred to by their numbers there. 
\section{Tight-binding model}
\label{sec:tb} 

We describe rhombohedral $N$-layer graphene by the continuum tight-binding model of Refs.~\cite{koshino2009trigonal,zhang2010band,xiao2011density, parra2025nematic,kumar2025superconductivity}.  In the basis $(A_1,B_1,A_2,B_2,\dots,A_N,B_N)$ and near the K valley, the Hamiltonian is $H=H_0+H_d+H_D$, where
\begin{equation}
  H_0=\begin{pmatrix}
    V_0 & V_N & V_{NN} & & \\
    V_N^{\dagger} & V_0 & V_N & \ddots & \\
    V_{NN}^{\dagger} & V_N^{\dagger} & V_0 & \ddots & V_{NN}\\
     & \ddots & \ddots & \ddots & V_N\\
     & & V_{NN}^{\dagger} & V_N^{\dagger} & V_0
  \end{pmatrix},
  \label{eq:H0}
\end{equation}
with the intralayer, nearest-layer and next-nearest-layer blocks
\begin{equation}
  V_0=-v_0\begin{pmatrix} 0 & \pi\\ \pi^{*} & 0\end{pmatrix},\qquad
  V_N=\begin{pmatrix} -v_4\pi & -v_3\pi^{*}\\ \gamma_1 & -v_4\pi\end{pmatrix},\qquad
  V_{NN}=\begin{pmatrix} 0 & \gamma_2/2\\ 0 & 0\end{pmatrix}.
  \label{eq:blocks}
\end{equation}
Here $\pi=k_x+ik_y$ is measured from K, $v_i=(\sqrt{3}/2)\,a\,\gamma_i$ with the lattice constant $a=0.246$~nm, and the hopping parameters are  ($\gamma_0$, $\gamma_1$, $\gamma_2$, $\gamma_3$, $\gamma_4$, $\delta$) = (3.1, 0.38, -0.015, -0.29, -0.141,0.0105) eVs. The dimer potential acts on the sites that belong to interlayer dimers,
\begin{equation}
  H_d=\delta\,\mathrm{diag}\big(0,\,-1,\;1,\,1,\;\dots,\;1,\,1,\;1,\,0\big),
  \label{eq:Hd}
\end{equation}
and the displacement field enters as a layer potential that falls linearly across the stack,
\begin{equation}
  H_D=\mathrm{diag}\big(U_1 I_2,\dots,U_N I_2\big),\qquad
  U_\ell=D\,\frac{N+1-2\ell}{N-1},\quad \ell=1,\dots,N,
  \label{eq:HD}
\end{equation}
with $I_2$ the $2\times2$ identity, so that the potential difference between the outer layers is $2D$, where  $D$ is the displacement-field parameter.

\begin{figure}[tb]
  \includegraphics[width=8.6cm]{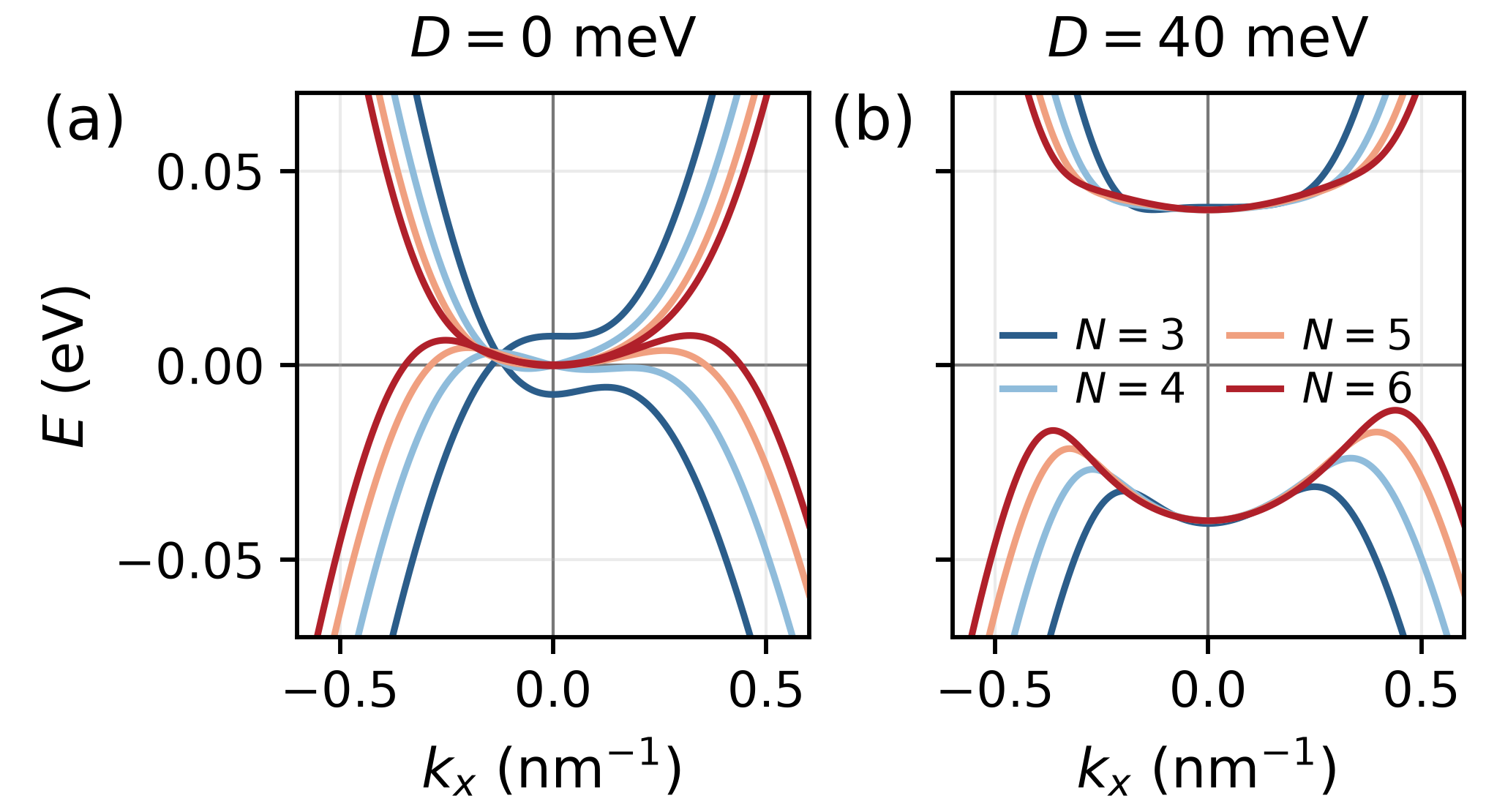}
  \caption{Low-energy bands of rhombohedral $N$-layer graphene along $k_y=0$ (defined with respect to K), from $H=H_0+H_d+H_D$, for $N=3-6$.  (a) At $D=0$ the conduction and valence bands touch at K.  (b) At $D=40$~meV a gap of $2D$ opens.}
  \label{fig:bands}
\end{figure}

\begin{figure}[tb]
  \includegraphics[width=8.6cm,height=8.6cm,keepaspectratio]{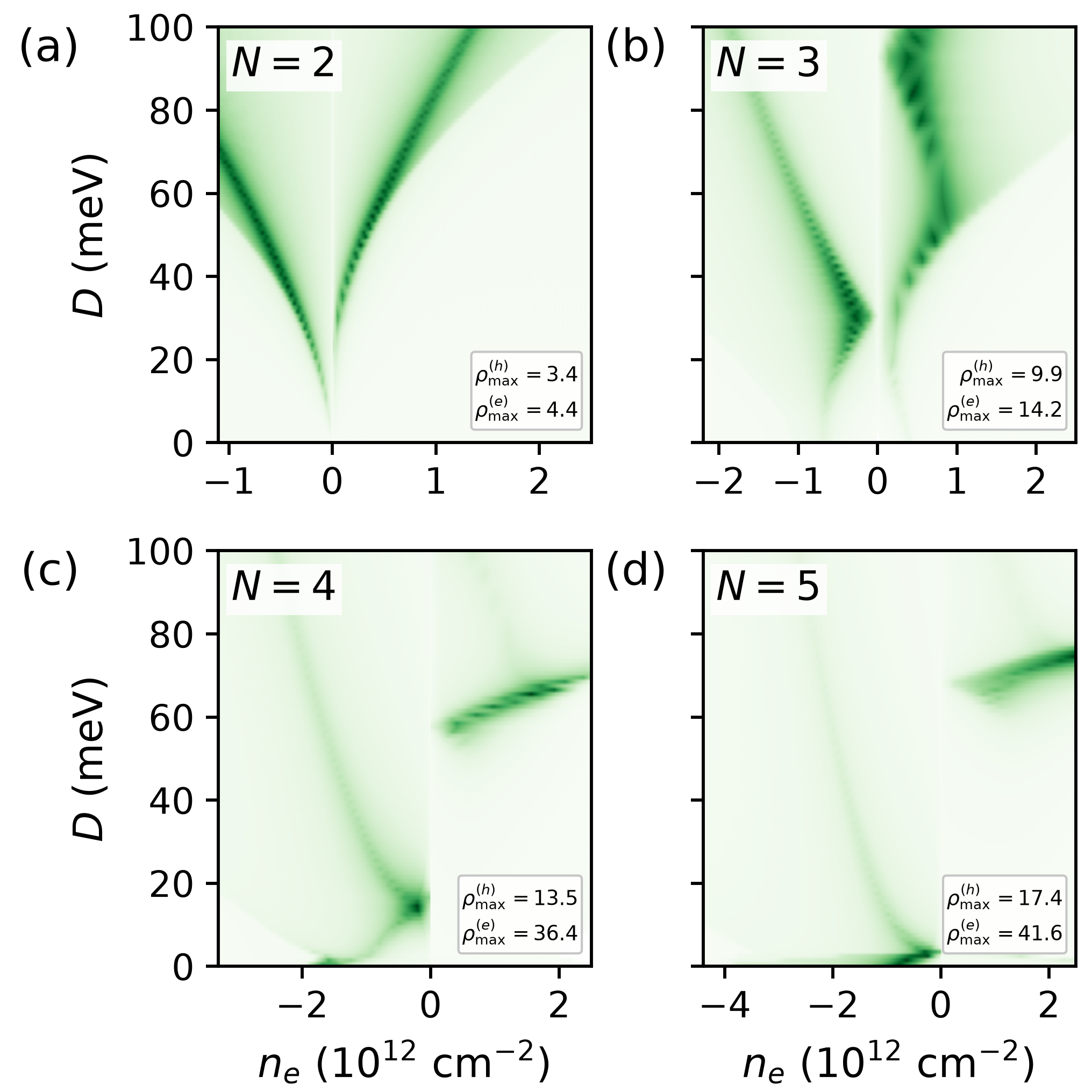}
  \caption{Density of states of the noninteracting model ($U=J=0$) for  rhombohedral graphene with (a) $N=2$, (b) $N=3$, (c) $N=4$, and (d) $N=5$ layers, in eV$^{-1}$nm$^{-2}$. }
  \label{fig:dos0}
\end{figure}

The eigenvalues of $H$ are computed on a $512\times512$ grid covering $|k_x|,|k_y|\le0.6$~nm$^{-1}$ around each valley.  The density of states per flavor is
\begin{equation}
  \rho_1(E)=\int\!\frac{d^2k}{(2\pi)^2}\sum_{b}\,\delta_\eta\big(E-\varepsilon_b(\bm k)\big),
  \label{eq:dos}
\end{equation}
and the carrier density per flavor measured from the band edge is $n_1(E)=\int_{E_c}^{E}\rho_1\,dE'$ on the electron side and $\int_{E}^{E_v}\rho_1\,dE'$ on the hole side, with $E_c$ ($E_v$) the conduction-band minimum (valence-band maximum).  
Figure~\ref{fig:bands} shows the computed low-energy bands and Fig.~\ref{fig:dos0}  the resulting noninteracting density of states (DoS).

\section{Flavor symmetry breaking}
\label{sec:stoner}

\begin{figure}[t]
\includegraphics[width=8.6cm,height=8.6cm,keepaspectratio]{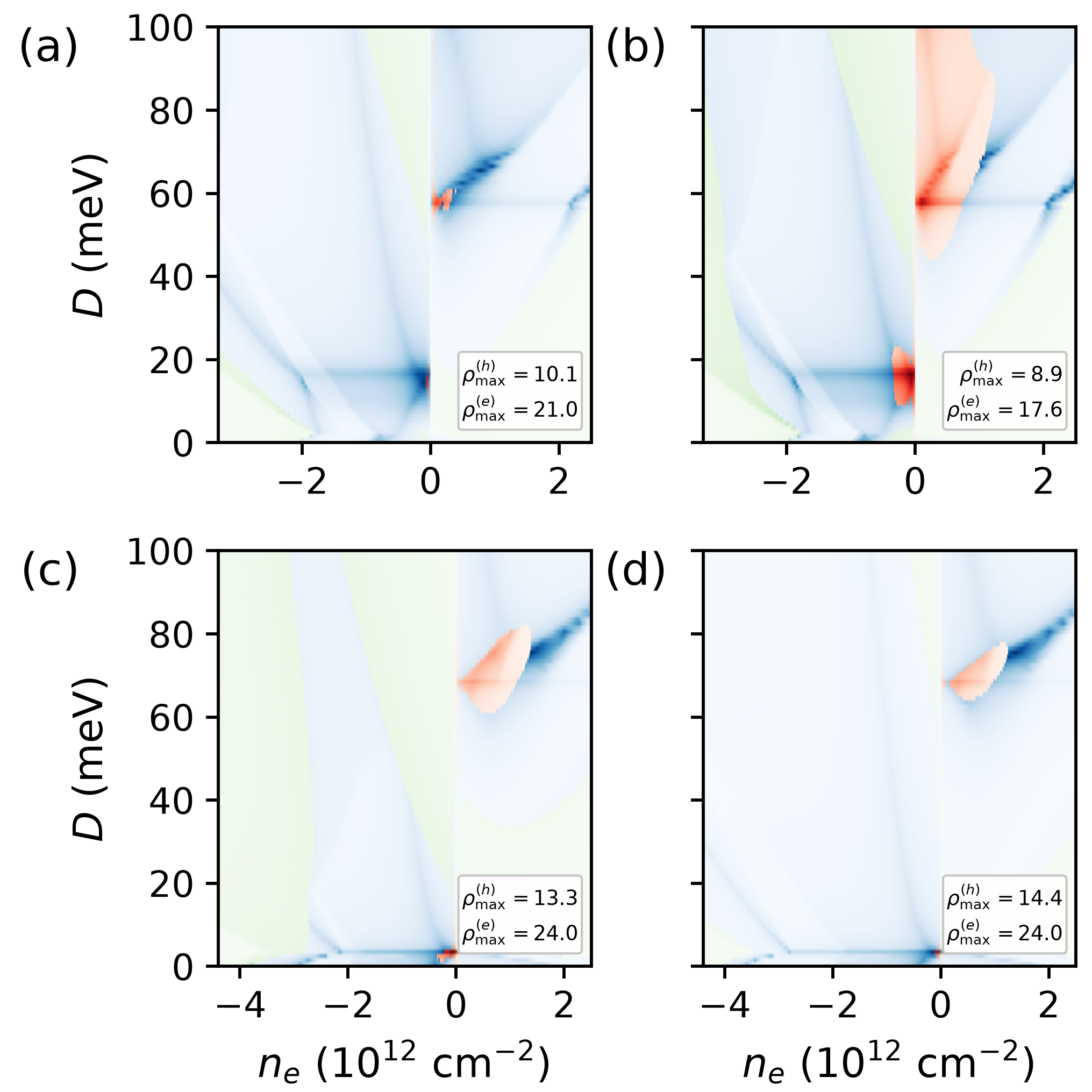}
\caption{Density of states in the rigid-band Stoner approximation for
  alternative interaction parameters: $N=4$ with (a) $(U,J)=(11,-16)$ and
  (b) $(15,-9)$~eV, and $N=5$ with (c) $(10,-9)$, and (d) $(13,-18)$~eV. Units are in $\rm{eV^{-1}nm^{-2}}$.}\label{FigA11}
\end{figure}

The interaction-driven symmetry breaking between the four spin-valley flavors is treated in the rigid-band Stoner approximation~\cite{zondiner2020cascade,zhou2021half,ghazaryan2023multilayer}: each flavor $\alpha$ fills the noninteracting band rigidly up to its own density $n_\alpha$, and the densities are chosen to minimize, at fixed total carrier density $n=\sum_\alpha n_\alpha$,
\begin{equation}
  E\big[\{n_\alpha\}\big]=\sum_{\alpha}\varepsilon(n_\alpha)
    +U A_{\rm uc}\sum_{\alpha<\beta}n_\alpha n_\beta + E_J ,
  \label{eq:stoner}
\end{equation}
 where $\varepsilon(n)$ is the kinetic energy of one flavor filled to density $n$, $A_{\rm uc}=(\sqrt3/2)a^2$ is the unit-cell area, and $U$ and $J$ are the Stoner and Hund parameters, respectively, in eV. For a given total carrier density $n$, the energy minimization determines the carrier distribution into majority ($n_1$) and minority ($n_2$) isospin flavors, associated with degeneracies $g$ and $4-g$, respectively. The effective interacting DoS, $\rho_{\mathrm{int}}$, is constructed by linearly combining the bare DoS evaluated at these flavor-resolved partial densities $\rho_{\mathrm{int}} = g \rho(n_1) + (4 - g) \rho(n_2)$, where $\rho(n_i)$ represents the noninteracting DoS mapped to the chemical potential that yields the partial density $n_i$. An alternative set of parameters ($U,J$) for $N = 4$ and $N = 5$ are shown in Fig.\ \ref{FigA11} as explained in the main text.

\begin{figure}[t]
\includegraphics[width=8.6cm,height=17.2cm]{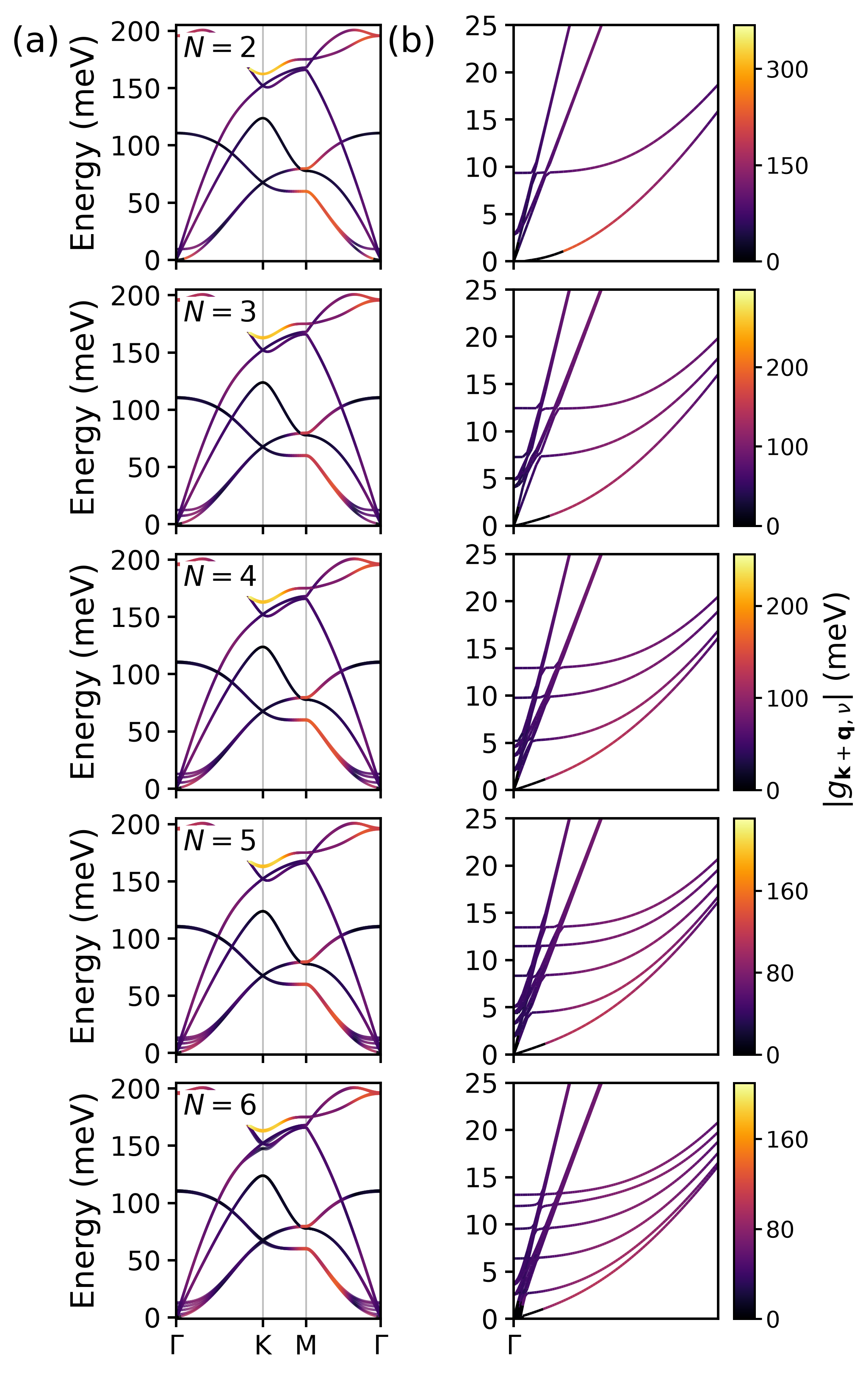}
\caption{First-principles phonon dispersions of multilayer rhombohedral graphene with $N=2 - 6$ layers. For each layer number, (a) shows the phonon dispersion along the high-symmetry path of the Brillouin zone, and (b) an enlarged view near the $\Gamma$ point. The color scale indicates the magnitude of the electron-phonon matrix elements $g_{\bm{k}+\bm{q},\nu}$.}\label{FigA4}
\end{figure}

\section{Density functional theory}
\label{sec:dft}

First-principles density functional theory (DFT) and density functional perturbation theory (DFPT) calculations were carried out using \textsc{Quantum ESPRESSO} \cite{giannozzi2009quantum,giannozzi2017advanced}. We employ the local density approximation (LDA), which has been extensively validated for phonons and electron-phonon interactions in graphene-based systems \cite{yan2008phonon,yan2009electron,park2008electron,borysenko2010first,borysenko2011electron,gao2024first}. Ground-state charge densities were computed on a $24 \times 24 \times 1$ Monkhorst-Pack $\bm{k}$-point grid, employing Marzari-Vanderbilt cold smearing with a broadening parameter of 0.02 Ry. The two-dimensional Coulomb truncation scheme, described in Ref.\ \cite{2DQE}, was used. Phonon dispersions and electron-phonon matrix elements were computed via DFPT on a $6 \times 6 \times 1$ $\bm{q}$-point mesh. Eliminating spurious imaginary phonon frequencies near the $\Gamma$ point requires a selfconsistency threshold of $10^{-14}$ to $10^{-15}$. Maximally localized Wannier functions (MLWFs) were constructed using the \textsc{Wannier90} code \cite{pizzi2020wannier90}. The electron-phonon matrix elements $g_{\bm q\nu}$ are interpolated onto a fine $128 \times 128 \times 1$ mesh using the \textsc{Perturbo} package \cite{zhou2021perturbo}, applying a soft phonon frequency cutoff of 1 meV to regularize the scattering rates.

The calculated phonon dispersions for $N=2 - 6$ are shown in Fig.\ \ref{FigA4}. The overall phonon spectrum changes little with increasing layer number, apart from the emergence of $N-1$ additional shear and breathing modes near the $\Gamma$ point. The color scale represents the magnitude of the electron-phonon matrix elements. A systematic reduction of the electron-phonon coupling with increasing layer number is evident. We attribute this reduction to enhanced electronic screening associated with the larger Fermi surface of thicker multilayers.

\section{Acoustic versus Optical Phonons}
\begin{figure}[t]
\includegraphics[width=8.6cm,height=4.3cm,keepaspectratio]{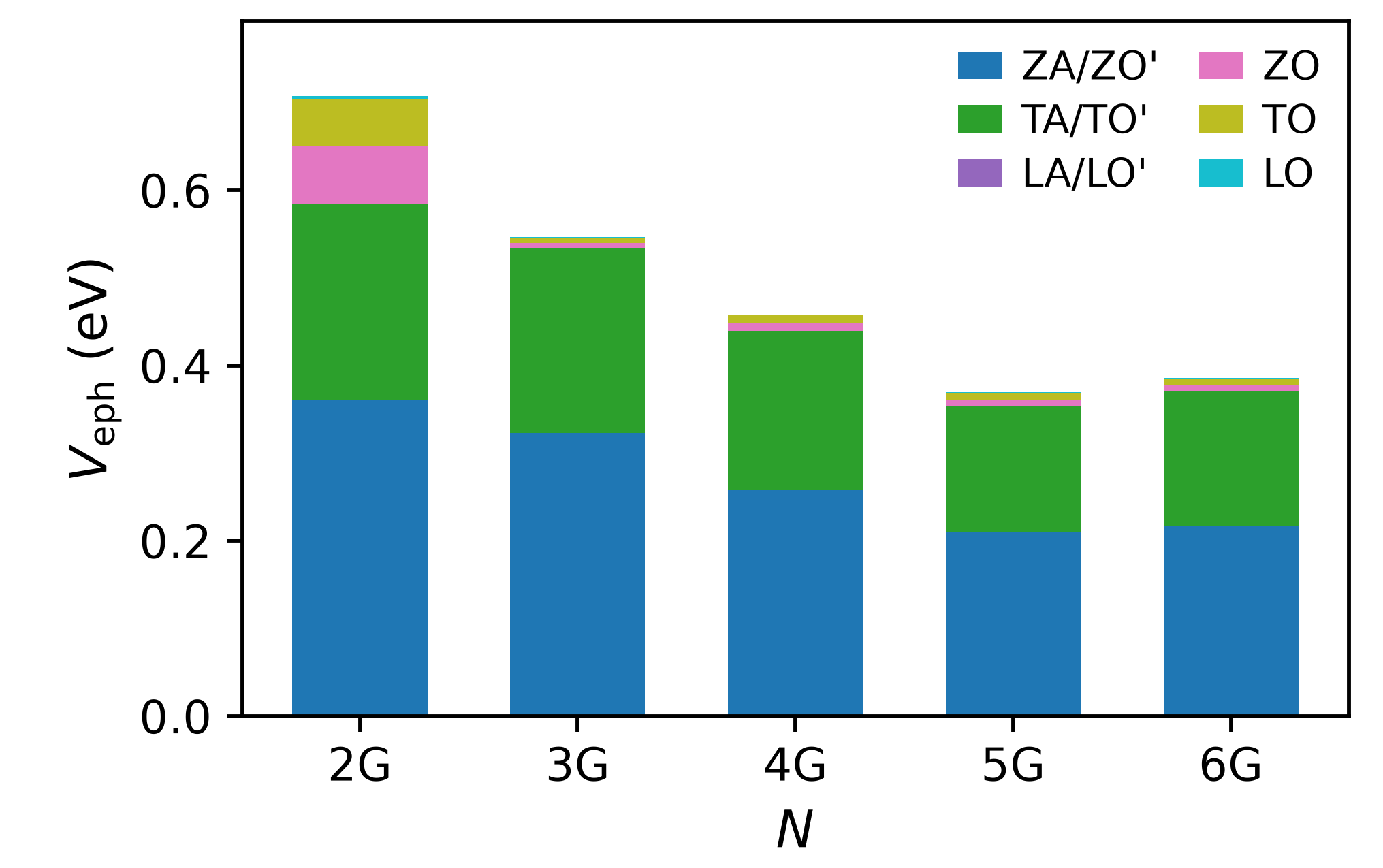}
\caption{Evolution of the electron-phonon coupling with the number of graphene layers for scattering of electrons near the $\Gamma$ point. The coupling is decomposed into (a) the total contribution and (b) contributions resolved by phonon branch.}\label{FigA2}
\end{figure}

In the main text we focus on $g(\bm{q},\bm{k}=K)$, since the electronic states at the Fermi level are centered around the K valleys. Figure \ref{FigA2} shows the corresponding calculation for $g(\bm{q},\bm{k}=\Gamma)$. In this case, the electron-phonon coupling is dominated by acoustic phonons, in contrast to the $g(\bm{q},\bm{k}={\rm K})$ results discussed in the main text. Despite this difference, retardation effects imply that superconductivity remains primarily mediated by acoustic phonons in both cases. However, the much stronger coupling obtained for $g(\bm{q},\bm{k}=\Gamma)$ would predict critical temperatures significantly higher than the experimentally observed $T<1$\,K.

\section{Migdal-Eliashberg Theory}
\label{sec:me}
We solve the Migdal-Eliashberg equations selfconsistently to determine the superconducting gap of the system,
\begin{eqnarray}
    Z(\bm{k},i\omega_n)&=&1+\frac{\pi T}{\omega_n}\sum_{\bm{k}^{\prime},m}\lambda(\bm{k}-\bm{k}',i\omega_n-i\omega_m)\frac{\omega_m Z(\bm{k}',i\omega_m)}{\Theta(\bm{k}',i\omega_m)},
    \\
    \phi(\bm{k},i\omega_n)&=&\pi T\sum_{\bm{k}',m}\lambda(\bm{k}-\bm{k}',i\omega_n-i\omega_m)\frac{\phi(\bm{k}',i\omega_m)}{\Theta(\bm{k}',i\omega_m)},
\end{eqnarray}
where $\Theta(\bm{k}',i\omega_m)=\sqrt{Z(\bm{k}',i\omega_m)^2+\phi(\bm{k}',i\omega_m)^2}$ and $\lambda(\bm{q},iv_m)=\sum_{\nu}\lambda^{E_{\rm F}}_{\bm{q}\nu}\frac{\omega^2_{q\nu}}{\omega^2_{\bm{q}\nu}+(\nu_m)^2}$.  $Z(\bm{k},i\omega_n )$ is the mass renormalization and $\phi (\bm{k},i\omega_n)$ the superconducting order parameter. The superconducting gap is defined as $\Delta = {\rm max}(\phi(\bm{k},i\omega_n)/Z(\bm{k},i\omega_n))$. The equations are  solved on the imaginary axis at fixed temperature, on the fermionic grid $\omega_n = (2n + 1)\pi T$. The cutoff of the Matsubara frequency is chosen as $\omega_c=1.05\max \omega_{\bm{q}n}$ where $\omega_{\bm{q}n}$ are the phonon frequencies determined from DFT+DFTP. As much as 25000 Matsubara frequencies are needed to cover the entire Matsubara grid.

\section{Intra- and inter-valley interactions}
In the calculations presented in the main text, we consider phonons with small momenta, corresponding to intra-valley scattering, with $|q_x|, |q_y| \leq 0.6~\mathrm{nm}^{-1}$. This choice is analogous to the momentum range considered for the electrons and allows us to capture the valley-polarized superconducting state corresponding to $\gamma = 2$. Below, we show that superconductivity also emerges when phonons with momenta spanning the entire Brillouin zone are included, thereby allowing for inter-valley scattering. In this case, however, the resulting $T_c$ is generally higher than the experimentally observed values.

For intra-valley pairing, the dominant phonon-mediated processes involve small momentum transfer. In this regime, the Coulomb interaction can induce Kohn–Luttinger superconductivity that may coexist with the phonon-mediated pairing \cite{you2022kohn,cea2022superconductivity,ghazaryan2021unconvnetional,li2023charge}. Consequently, introducing an Anderson–Morel pseudopotential $\mu^*$ as an repulsive interaction in the intra-valley calculation is not physically well motivated, since it would partially suppress a pairing channel that can itself arise from the Coulomb interaction.

Here, instead, we use the Coulomb pseudopotential $\mu^*$ to model the inter-valley case, for which its inclusion is physically motivated. In this case, the initially large superconducting gaps are substantially renormalized by the Coulomb repulsion, yielding gap magnitudes and transition temperatures that are 
consistent with the experimentally relevant energy scales.

\begin{figure*}[t]
  \centering
  \includegraphics[width=\linewidth]{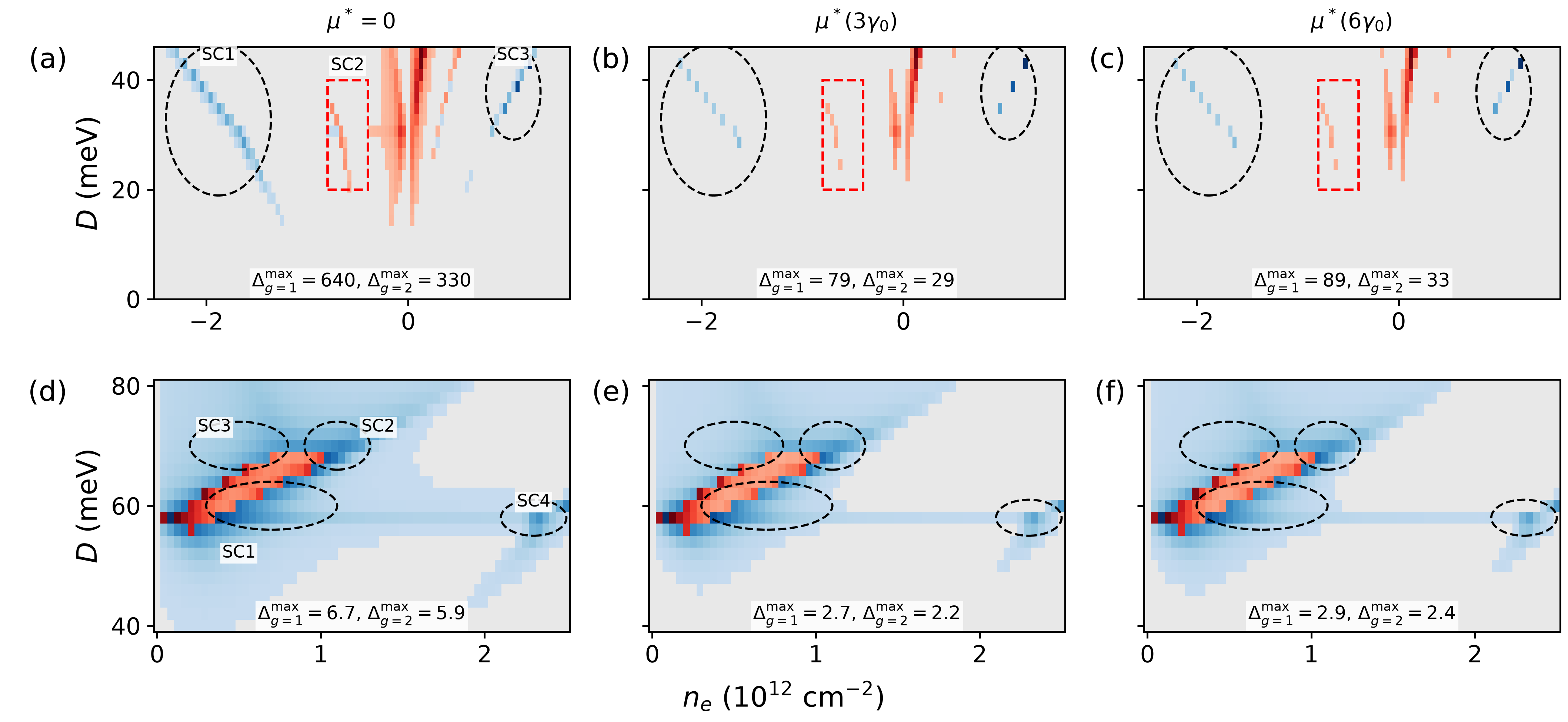}
  \caption{Calculated superconducting gap over the  $(D,n_e)$  plane for the inter-valley interaction, with and without the Coulomb pseudopotential. Rows: the trilayer at $T = 25$ mK (top) and the tetralayer at $T = 75$ mK (bottom). Columns: $\mu^*$ = 0, then $\mu^*$ evaluated with $E_{\rm cut} = 3\gamma_0$ (9.3 eV) and $6\gamma_0$  (18.6 eV).  The legend gives the maximum gap in $\mu$eV for the trilayer and meV for the tetralayer.}
  \label{FigA6}
\end{figure*}

\section{Coulomb pseudopotential $\mu^*$}
\label{sec:mustar}
We proceed to calculate the $\mu^*$ from many-body interaction instead of fitting the value to experiment. Following \cite{Schonhoff2016}, the Coulomb coupling and the pseudopotential are defined as 
\begin{eqnarray}
    \mu = N(E_{\rm F})\langle \langle |\Lambda_{\bm{kk'}}|^2 W(\bm{k}-\bm{k}') \rangle \rangle_{\rm FS},\quad \mu^*=\frac{\mu}{1+\mu \ln{(E_{\rm cut}/\omega_{\rm log})}} ,
\end{eqnarray}
where $\Lambda_{\bm{kk}'}=\braket{u_{\bm{k}}|u_{\bm{k}'}}$. The
interaction is the RPA-screened, gate-screened Coulomb
potential~\cite{ghazaryan2021unconvnetional}
\begin{equation}
  W(\bm{q})=\frac{V_0(q)}{1+\Pi(\bm{q})V_0(q)},\qquad
  V_0(q)=\frac{2\pi e^2}{\epsilon q}\tanh(qd),
  \label{eq:W}
\end{equation}
with $\epsilon=16$ for trilayer graphene, $\epsilon = 20$ for tetralayer graphene and gate distance $d=20$~nm \cite{zhang2025layer}. For the interlay repulsion we take $W({\rm K}-{\rm K}')$.  The polarization of one flavor is the static Lindhard function of the band that crosses the Fermi level,
\begin{equation}
    \Pi_{\alpha}(\bm{q}) = \frac{1}{N_{\bm k}}\sum_{\bm k}\,
  \big|\Lambda_{\bm k,\bm k+\bm q}\big|^{2}\,
  \frac{f(\varepsilon_{\bm k})-f(\varepsilon_{\bm k+\bm q})}
       {\varepsilon_{\bm k+\bm q}-\varepsilon_{\bm k}}.
  \label{eq:lindhard}
\end{equation}  
The total polarization sums the four flavors at
their own Fermi levels,
\begin{equation}
  \Pi(\bm{q}) = g\,\Pi_\alpha\big(\bm{q};E_{\rm F}^{\rm maj}\big) + (4-g)\,\Pi_\alpha\big(\bm{q};E_{\rm F}^{\rm min}\big).
  \label{eq:pisum}
\end{equation}
The logarithmic phonon frequency is
\begin{equation}
  \omega_{\log} = \exp\!\left[
    \frac{\sum_{\bm q\nu}\lambda^{0}_{\bm q\nu}\,\ln\omega_{\bm q\nu}}
         {\sum_{\bm q\nu}\lambda^{0}_{\bm q\nu}}\right].
  \label{eq:wlog}
\end{equation}
The expression for $\phi$ of the Migdal-Eliashberg calculations is then rewritten as
\begin{eqnarray}
    \phi_n=\pi T\sum_m [\lambda(i\omega_n-i\omega_m)-\mu^*]\frac{\phi(\bm{k}',i\omega_m)}{\Theta(\bm{k}',i\omega_m)}.
\end{eqnarray}
We present the numerical solution of the Migdal-Eliashberg calculation in Fig.\ \ref{FigA6} for two values of cutoff $E_{\rm cut}$.  We find that superconductivity generally survives in the presence of $\mu^*$ and for different cutoffs.

\end{document}